# Research and Simulation on Drivers' Route Choice Behavior Cognition Model

Na Lin, Hong-Dong Liu, Chang-Qing Gong

School of Computer, Shenyang Aerospace University
Shenyang, Liaoning 110136, China

**Abstract**
This paper studied the behavior-cognitive model of drivers during their travel based on the current research on driver behavior. Firstly, a route choice behavior-cognitive model was proposed for describing the decision-making mechanism of drivers during his travel; then, simulation experiments were carried out on the co-simulation VBc-vissim platform. From the experimental results, dynamic behavior features of drivers during their travel can be properly explained by the behavior-cognitive model, thus optimal path can be obtained from this model.

## 1. Introduction

Intelligent traffic system is an interactive and stochastic system among driver, vehicle and outer environment. As the core of this system, driver, which has unique thought and can modify his behavior in accordance with former experience, is the only subjective factor. During the whole traveling process, a driver can continuously change his behavior with reference to destination, traffic report and so on. It is pointed by traffic stream theory that the macro traffic stream phenomenon can be reproduced through modeling of drivers' individual levels. Through which the actual status of traffic stream under different circumstances can be reproduces and drivers' personal interests can also be reflected. This paper concludes and divides the previous works into three aspects.

The first aspect concerns the research of drivers' route choice behavior based on SP survey of VMS information. A Stated Preference (SP) survey for drivers' route choice behavior under Variable Message Sign (VMS) was carried out in [1] on the basis of analysis of drivers' route choice factors and a BP neural network model was established under VMS. In according to the survey data, dynamic information and drivers' route selection tendency based logistic regression model is proposed in [2]. By using VMS and by collecting data through SP, a dynamic route choice behavior model which can provide traveling time, traffic situation information was proposed in [3] after discrete analysis of these selected data. The influences from information and experience to drivers were discussed through SP survey in [4] and a discrete route choice model was proposed for verifying the influences. While in [5], the macro traffic stream data obtained from microwave testing and information from VMS were combined for analyzing drivers' route choice behavior.

The second aspect concerns multi-agent technology. Agent model is a very powerful simulation technology. Agent model based traffic simulation techniques, advantages and existed problems were briefly introduced in [6]. A dynamic route choice agent negotiation model was introduced in [7] within which drivers, information publisher and system manager were all treated as separate agent unit for modeling. And in [8], behaviors of an individual driver under information influence were simulated through an agent based mental model.

And the third aspect concerns the influence from drivers' emotions to route choice. The resulted influences to driving mode or drivers' behaviors were demonstrated in [9] through an emotion-cognitive model. Another cognitive simulation model was presented in [10] after a psychoreaction analysis of drivers.

## 2. Our Work

### 2.1 Research Scheme

The research of modern traffic system requires combining factors, such as drivers, vehicles, roads and surrounding environment, into a static-dynamic integrated system which has the non-linear, random and time varying features. During the driving process, drivers are accompanied with complicated cognitive variation. Only by establishing a more proper route choice behavior model through fully considering drivers' route choice behavior features, can we provide actual theory model for micro-traffic stream simulation [11]. This paper mainly focuses on the influences from drivers' behavior cognition to route choice and a feasible traveling model is proposed.

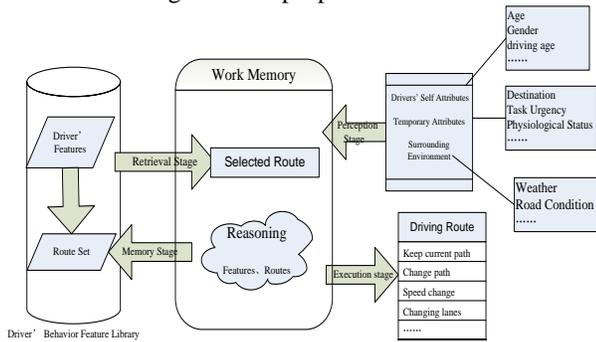

Fig.1 Drivers route choice behavior cognition process

Drivers' route choice behavior cognition process is given in Fig.1. The establishment of drivers' feature library can make route selection more convenient and avoid repeat computation. The library contains corresponding relations. For each set of behavior features, there is a corresponding route choice. The process is comprised by the following stages:

Perception stage: Drivers' self attributes (age, gender, driving age, etc.), temporary attributes (destination, task urgency, physiological status, etc.) and surrounding environment (weather, road condition, etc.) can be sensed by work memory;

Retrieval stage: Search the feature library in work memory for corresponding route choice.

Execution stage: Drive into the road obtained in retrieval stage. Periodically sense the variation of temporary attributes and surrounding environment and repeat the retrieval process.

Reasoning stage: If the feature library does not contain the current feature set, a reasoning process would occur according to current feature and selected road. The newly generated corresponding relation would be stored in the feature library.

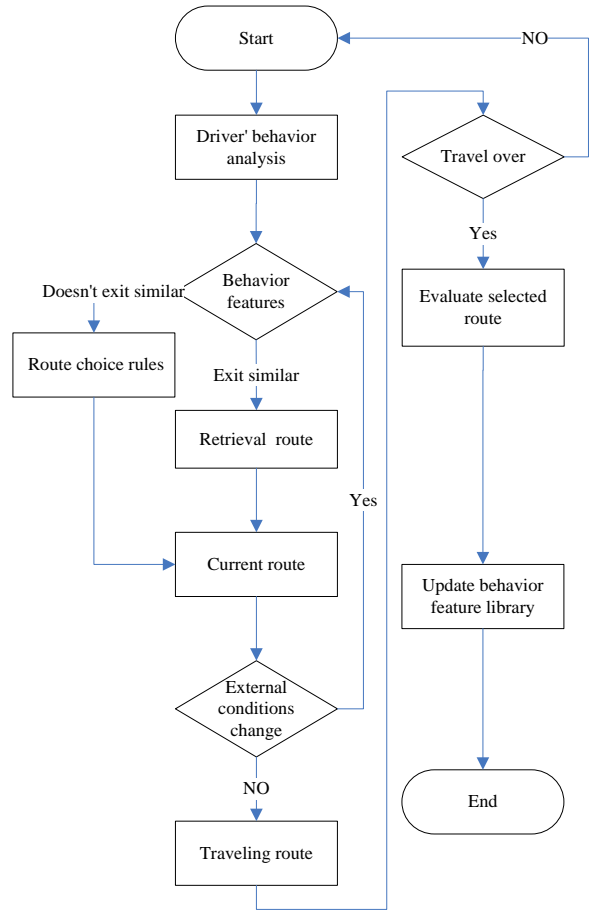

Fig.2 Drivers' traveling process

The drivers' detailed traveling process is as Fig.2 shows:
(1) Before traveling, driver's behavior features are analyzed. If his feature set can be retrieved from feature library, the traveling route will be obtained; or the route will be re-computed according to route choice rules;
(2) During traveling, if variation of temporary attributes and surrounding environment is sensed, repeat (1);
(3) After traveling, the traveling information of this time is evaluated. If the feature library contains current driver feature set, select and store the better route choice according to evaluation result; or, store the new generated corresponding relation into feature library.

2.2 Route choice rules

For the can-not-find situations in the above route choice cognition process, the route is selected according to the rules in Fig. 3.

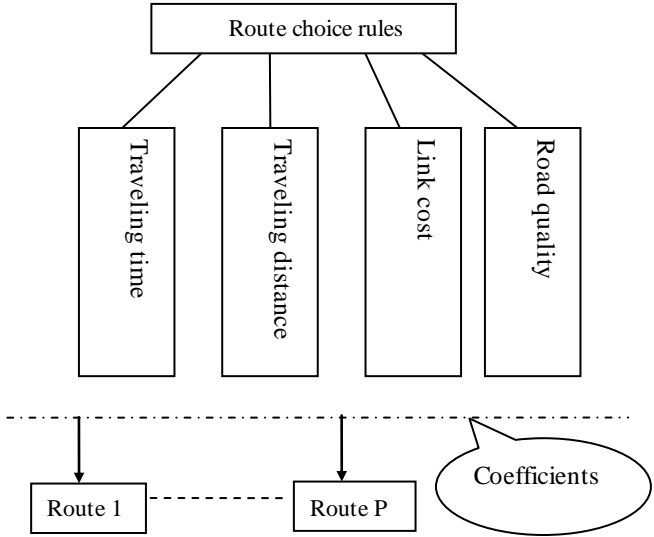

Fig. 3 Route choice rules

Travel Time is not the only factor to influence route choice. There are at least two other major influences: travel distance and financial cost (e.g. tolls). In contrast to travel times these factors are not depending on the traffic situation and thus have not to be determined by simulation. To cover all three major influences on route choice, for each of the edges in the network the so called general cost is computed as a weighted sum:

$$\text{general cost} = \alpha \times \text{travel time} + \beta \times \text{travel distance} + \gamma \times \text{link cost} + \delta \times \text{road quality} + \sum \text{supplement2} \quad (1)$$

The coefficients α, β, γ and δ can be defined by the user. In VISSIM the weights are specific to vehicle types and allow the modeling of driver groups with different route choice behavior.

The travel distances are determined by the geometry of the links. The financial cost of an edge is the sum of the costs of all links that are contained in that edge. The individual cost of a link is computed by multiplying the traveled distance on that link by the cost specified as a link attribute plus adding supplement1.Supplement2 is an additional cost value per link that is not weighted by the factor γ but just added into the general cost of the edge.

2.3 Model for optimizing

A route is a sequence of edges that describes a path through the network. Since normally more than one route exists between an origin and a destination in the network VISSIM has to model the driver's decision which route to take. For the beginning let's assume that the set of available routes is known for a certain origin-destination pair. The route choice is a special case of the discrete choice problem. For a given set of discrete alternatives the probabilities for the alternatives to be chosen must be determined. For traffic assignment we need to define a utility function to assess each route in the set and a decision function based on this assessment.

The general cost for a route is then simply defined as the sum of the general costs of all its edges:

$$C_R = \sum_{a \in R} C_a \quad (2)$$

Where, C=general cost;
R=a route; a=an edge belonging to R.

One basic assumption in VISSIM's route choice model is that not all drivers use the best route but all routes available can be used. Of course more traffic should be assigned to "better" routes than to "worse" routes. To assess how "good" a route is, we use the general cost of the route as explained in the section above. The general cost is obviously the inverse of what is called a utility value in discrete choice modeling. So we use as an utility function the reciprocal of the general cost:

$$U_j = \frac{1}{C_j} \quad (3)$$

Where, $U_j$=utility of route j; $C_j$=general cost of route j.

The most widely used and thus theoretically best analyzed function to model discrete choice behavior is the Logit function:

$$p(R_j) = \frac{e^{\mu U_j}}{\sum_i e^{\mu U_i}} \quad (4)$$

Where, $U_j$=utility of route j; $p(R_j)$=probability of route j to be chosen; $\mu$ =sensitivity factor of the model(>0).

The sensitivity factor determines how much the distribution reacts to differences in the utilities. A very low factor would lead to a rather equal distribution with nearly no regard of utility, and a very high factor would force all drivers to choose the best route. If we use the Logit function with an utility function defined as above, we end up with the situation that the model considers the difference between 5 and 10 minutes of travel time to be the same as the difference between 105 and 110 minutes of travel time, since the Logit function is invariant against translation and considers only the absolute difference of the utilities. Obviously that is not appropriate for deciding route choice, since in the real world two routes having travel times of 105 and 110 minutes would be considered nearly the same, whereas 5 and 10 minutes are much of a difference. The solution adopted in VISSIM is to use the so called Kirchhoff distribution formula:

$$p(R_j) = \frac{U_j^k}{\sum_i U_i^k} \quad (5)$$

Where, Uj=utility of route j; p(Rj)=probability of route j to be chosen; k=sensitivity of the model.

Again the sensitivity k in the exponent determines how much influence the differences in utility have. In this formula, the relative difference in utility determines the distribution, so that we will see only a small difference between the 105 and 110 minute routes, whereas the 5 minute route will receive much more volume than the 10 minute route. Actually the Kirchhoff distribution formula can be expressed as a Logit function, if the utility function is transformed to be logarithmic:

$$p(R_j) = \frac{U_j^k}{\sum_i U_i^k} = \frac{e^{k \log U_j}}{\sum_i e^{k \log U_i}}$$
$$= \frac{e^{-k \log C_j}}{\sum_i e^{-k \log C_i}} \quad (6)$$

Where, Cj is the general cost of route j.

## 3. Simulation and Analysis

### 3.1 Simulation Frame

Fig. 4 shows the frame of both com interface of Visual Basic (VBc) and micro simulation platform vissim. VBc is in charge of drivers' decision behavior under external information before traveling (traveling route choice which determines the distribution of traveling). vissim is in charge of drivers' actual traveling status in road network after receiving origin-destination requirements from VBc. vissim is also for evaluating the performance of selected route and for computing new route according to road network variations.

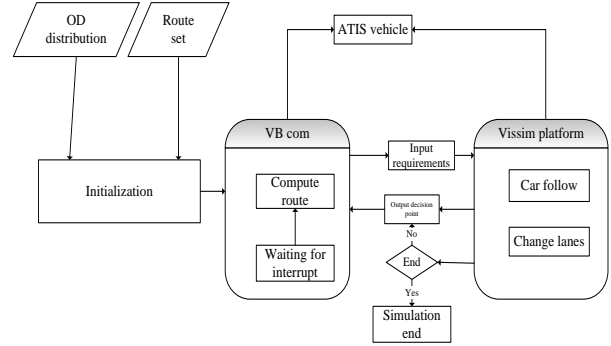

Fig. 4 VBc and vissim simulation platform

### 3.2 Road network model

Fig. 5 is an abstract road network which uses the map of Beijing city as simulation background images for vissim. Within Fig. 5, triangles represent nodes that lie at the edge of road network; quadrangles represent inner nodes; pentacles represent zones, lines represent roads and each zone has a parking lot. This road network contains 12 zones (z1-z12), 12 edge nodes and 12 inner nodes (1-24), 3 city express with bidirectional 5 lanes (Express 1-3), 1 major road with bidirectional 4 lanes (Major1), 2 minor roads with bidirectional 3 lanes (Minor1-2) and 1 slip bidirectional 2 lanes (slip1).

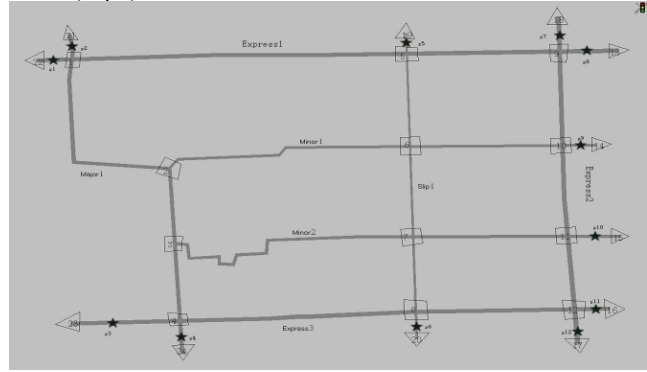

Fig. 5 Abstract road network

### 3.3 Simulation Experiment

Experiment simulates the traffic flow in flat time and in peek time separately. As will be seen, the road network cost is decreased by adapting the behavior cognition model. Experimental results are through comparisons of average travel costs. Experiment steps are as follows:

(1) The behavior before traveling is firstly simulated. The OD traffic flow distribution of each zone in flat time is resided in a .FMA file (as is shown in Tab.1).
(2) After running the model 1000 simulation clocks without road resistance, the OD traffic distribution file is imported to vissim, and the work period is set to 120 seconds.
(3) After each work period, the generated road network parameters are imported to VBc for a second computation. The behavior cognition model is applied during the computation and the newly generated OD traffic distribution file is imported to vissim again.
(4) Repeat step 2 and 3 until reaching a convergence.

Tab.1: Traffic OD distribution

| Zone No. | Z1 | Z2 | Z3 | Z4 | Z5 | Z6 | Z7 | Z8 | Z9 | Z10 | Z11 | Z12 |
|---|---|---|---|---|---|---|---|---|---|---|---|---|
| Z1 | 0 | 200 | 182 | 221 | 235 | 120 | 80 | 60 | 105 | 89 | 800 | 253 |
| Z2 | 198 | 0 | 25 | 32 | 19 | 160 | 120 | 4 | 19 | 52 | 4 | 30 |
| Z3 | 181 | 25 | 0 | 305 | 29 | 209 | 120 | 35 | 19 | 30 | 13 | 5 |
| Z4 | 210 | 31 | 300 | 0 | 80 | 240 | 131 | 45 | 30 | 130 | 85 | 190 |
| Z5 | 255 | 20 | 30 | 79 | 0 | 1398 | 19 | 107 | 30 | 228 | 333 | 305 |
| Z6 | 118 | 159 | 210 | 239 | 1397 | 0 | 330 | 296 | 225 | 330 | 322 | 15 |
| Z7 | 76 | 110 | 121 | 129 | 19 | 330 | 0 | 80 | 153 | 220 | 2 | 16 |
| Z8 | 63 | 5 | 34 | 49 | 106 | 297 | 80 | 0 | 280 | 17 | 14 | 32 |
| Z8 | 110 | 21 | 21 | 28 | 30 | 225 | 150 | 276 | 0 | 145 | 13 | 56 |
| Z10 | 92 | 49 | 30 | 132 | 229 | 315 | 221 | 16 | 144 | 0 | 68 | 97 |
| Z11 | 786 | 3 | 13 | 90 | 335 | 325 | 2 | 14 | 12 | 67 | 0 | 145 |
| Z12 | 260 | 28 | 6 | 180 | 205 | 15 | 16 | 32 | 56 | 98 | 146 | 0 |

3.4 Result analysis

A、Flat time traffic

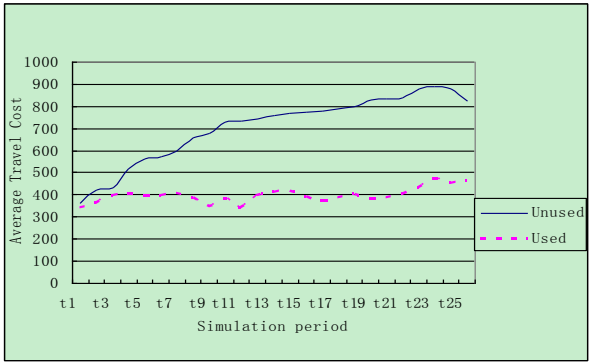

Fig.6 Evolution of average travel cost in flat time

B、Peek time traffic

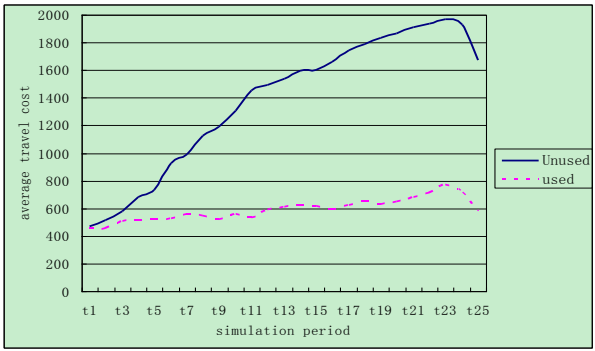

Fig.7 Evolution of average travel cost in peek time

Fig. 6 and Fig. 7 are about the average travel costs in flat time and in peek time separately. The result shows that the mean value and variance with behavior cognition model are all lower than those without. This means that the road network performance is greatly improved after adapting the cognition model.

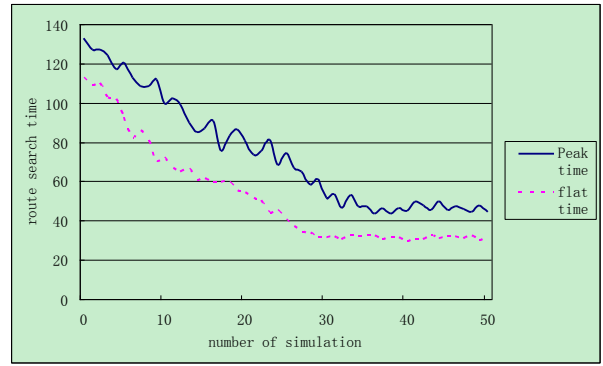

Fig. 8 Time curves of route search

Fig. 8 shows the time curves of route search in flat time and in peek time when the simulation model runs 50 times. After about 30 times, the curve for flat traffic reaches a stable route search time; for peek time traffic, simulation needs about 40 times of model running. It is also can be seen that route choice is apparently affected by drivers' behavior features after certain time of model running. Simulation in peek time requires more time, this is because the time required for route computing and drivers' behavior feature library querying in peek time is more than in flat time.

## 4. Conclusions

The intelligent traffic system has complex, dynamic and discrete features which make it hard to describe drivers' behaviors by a rigorous mathematical model. A driver's behavior cognition model is proposed in this paper which analyzed the evolution of drivers' cognition process. A road network is abstracted and tested on the VBc and vissim integrated simulation platform for modeling and execution. Drivers' road choice behaviors are better described by the model which provides a new idea for analyzing drivers' behaviors. Due to limitations of specific road network of different cities, the obtained result cannot merely be applied to other city road networks. Further, if the signal light intervals at corresponding crosses can be adjusted, the simulation speed would be further improved.

**Acknowledgments**

The Project is sponsored by Natural Science Foundation of Liaoning Province (No.20102175), "Liaoning Bai-Qian-Wan Talents Program" (No.2009921089, 2010921080), Liaoning Province Education Department Research Projects (No.L2010423), Graduate Education of Liaoning Province Innovation Projects and Program for Liaoning Excellent Talents in University.

**Na Lin** female, Shenyang, associate professor, master tutor, Northeastern University Postdoctoral. Research areas include Intelligent Transportation and Next Generation Network.

**Hong-dong Liu** male, master graduate student, research area is dynamic route guidance in the Intelligent Transportation.

**Chang-Qing Gong** male, Shenyang, professor, master tutor, Research areas include Quantum Computing and Internet of things.